\documentclass[doublecol]{epl2}

\usepackage{amsmath,mathrsfs}
\usepackage[dvipsnames]{xcolor}

\newcommand{\prt}{\partial}

\title{Plasma slab expansion into vacuum }
\shorttitle{Plasma slab expansion} 

\author{S. K. Ivanov \inst{1,2} \and A. M. Kamchatnov\inst{1,2}
} \shortauthor{S. K. Ivanov,  A. M. Kamchatnov}

\institute{
	\inst{1} Institute of Spectroscopy,
	Russian Academy of Sciences, Troitsk, Moscow, 108840, Russia\\
	\inst{2} Moscow Institute of Physics and Technology, Institutsky
	lane 9, Dolgoprudny, Moscow region, 141701, Russia
}


\abstract{
	The problem of collisionless plasma slab expansion into vacuum is solved within a
	two-temperature hydrodynamic approximation in the dispersionless limit of zero
	Debye radius. In framework of such an approach, the solution by the Riemann method
	provides quite accurate description of the whole process of plasma dynamics. It is
	shown that the dispersionless approximation agrees very well with
	exact numerical solution of the full system of plasma hydrodynamic equations.}

\begin{document}
	
	\maketitle
	
	\section{Introduction}
	
	Expansion of matter into vacuum is one of the canonical problems in fluid dynamics.
	In framework of plasma physics, it was studied first in Refs.~\cite{gpp-66,gpp-73}
	with the use of collisionless kinetic equation for slow motion of ions and under
	conditions of the thermal equilibrium for high-temperature electrons. This model
	was later modified in different directions and found applications to explanation
	of experiments on dynamics of plasma produced by interactions of very intensive
	laser pulses with matter (see, e.g., review article \cite{abp-13} and references in).
	As was shown in Refs.~\cite{gpp-66,gpp-73} (see also \cite{mp-79}), self-similar
	expansion of the plasma which occupies initially the half-space converge very
	fast to the dynamics of cold ions, so in the leading approximation one can
	neglect dispersive effects of finite Debye radius and use the purely hydrodynamic
	approach for quasi-neutral plasma (see, e.g., \cite{LL-10}).
	
	If a thin foil is irradiated by the laser pulse, then the plasma dynamics is not
	self-similar anymore, and the problem becomes much more complicated for the
	analytical approach, so it was studied mainly numerically (see, e.g.,
	\cite{bnbtb-04}). However, recently it has been shown that the classical
	Riemann method \cite{riemann,CH-89,Som64} can be successfully applied to this kind
	of problems. In particular, the problem of expansion of Bose-Einstein
	condensate released from a box-like trap was solved in Ref.~\cite{ik-19} and a
	detailed study of the Landau-Khalatnikov problem on expansion of high-temperature
	hadronic matter was given in \cite{kamch-19}. In this Letter we solve by this
	method the problem of expansion of the plasma slab in the hydrodynamic approximation.
	The solution provides the main characteristics of the bulk of ions in the
	expanding plasma.
	
	\section{Formulation of the problem}
	
	We assume that initially plasma occupies a slab $-l\leq x\leq l$ with a uniform
	density $n_0$ of ions whose temperature $T_i$ is much smaller than the temperature
	$T_e$ of electrons ($T_e \gg T_i$), so their thermal motion can be neglected. As was indicated in
	Introduction, in the process of plasma acceleration, as was shown in
	\cite{gpp-66,gpp-73}, the velocity distribution function degenerates exponentially
	fast to the $\delta$-function corresponding to the hydrodynamic approximation.
	Therefore one can apply to this problem the standard system of equations
	(see, e.g., \cite{LL-10})
	\begin{equation}\label{PlasmaDimention}
	\begin{split}
	& n_T+(nv)_X = 0, \\
	& v_T + v v_X + \frac{e}{m_i} \phi_X = 0, \\
	& \phi_{XX} = 4 \pi e \left( n_0 \exp{\frac{e\phi}{T_e}} - n \right),
	\end{split}
	\end{equation}
	where $n$ is the ion density, $v$ is the hydrodynamic velocity of
	the ions, $m_i$ is their mass, $\phi$ is the electric potential, $e$
	is the charge of electrons, and for simplicity we assume the charge of ions
	being equal to $-e$. Such a plasma is characterized by the Debye radius
	$D = (T_e/4\pi e^2n_0)^{1/2}$, ion plasma frequency
	$\Omega = (4\pi e^2n_0/m_i)^{1/2}$, and the ion-sound velocity
	$c_s=(T_e/m_i)^{1/2}$. This permits one to transform the system (\ref{PlasmaDimention})
	to convenient non-dimensional variables $t=\Omega T$, $x=D^{-1} X$,
	$u=v/c_s$, $\varphi = (e/T_e)\phi$ and to obtain
	\begin{equation} \label{PlasmaEq}
	\begin{split}
	& \rho_t+(\rho u)_x = 0, \\
	& u_t + u u_x +  \varphi_x = 0, \\
	& \varphi_{xx} =  e^{\varphi} - \rho.
	\end{split}
	\end{equation}
	This system is still too complicated for analytical treatment and it will be
	solved later numerically. However, if the initial width of the slab is much
	greater than the Debye radius, then the dispersive effects are relatively small
	and we can neglect the second order derivative in the last (Poisson) equation
	in the system (\ref{PlasmaEq}) and replace it by the Boltzmann equation
	$\rho=e^{\varphi}$. Then elimination of $\varphi$ yields the purely
	hydrodynamic system
	\begin{equation} \label{hydro}
	\begin{split}
	& \rho_t+(\rho u)_x = 0, \\
	& u_t + u u_x +  \frac{\rho_x}{\rho} = 0,
	\end{split}
	\end{equation}
	which should be solved with the slab initial distribution of the density
	\begin{equation} \label{distr}
	\begin{split}
	\rho(x,t=0) =
	\begin{cases}
	1, & \quad \text{if} \quad |x| \le l, \\
	0, & \quad \text{if} \quad |x| > l,
	\end{cases}
	\end{split}
	\end{equation}
	and $u(x,t=0) = 0$.
	
	The variables $\rho$ and $u$ have clear physical meaning, however, in physics of
	nonlinear waves other variables called {\it Riemann invariants} are more
	convenient (see, e.g., \cite{LL-6}), because in unidirectional wave propagation
	one of them is constant. In the case of the system (\ref{hydro}) equivalent to
	equation of gas dynamics with isothermal equation of state
	the Riemann invariants are calculated in standard way and they can be written as
	\begin{eqnarray} \label{t5-20.3}
	r_{\pm}=u\pm \int_0^{\rho}\frac{d\rho}{\rho}=u\pm \ln\rho,
	\end{eqnarray}
	where we used the fact that in our units the sound velocity is constant: $c_s=1$.
	The hydrodynamic equations (\ref{hydro}) transformed to these variables take
	a simple symmetric form
	\begin{equation}\label{t5-20.4}
	\frac{\prt r_{\pm}}{\prt t}+v_{\pm}\frac{\prt r_{\pm}}{\prt x}=0,
	\end{equation}
	where the characteristic velocities are equal to
	\begin{equation}\label{t5-20.5}
	v_{\pm}=u\pm 1=\frac12(r_++r_-)\pm 1.
	\end{equation}
	If the Riemann invariants are found, then the physical variables are expressed in term of them
	by the formulas
	\begin{equation}\label{t5-22.16}
	\begin{split}
	\rho(x,t)&=\exp\left(\frac{r_+(x,t)-r_-(x,t)}2\right),\\
	u(x,t)&=\frac12[r_+(x,t)+r_-(x,t)].
	\end{split}
	\end{equation}
	Now we can turn to solving the formulated problem.
	
	\section{Rarefaction waves}
	
	The evolution starts with formation of two rarefaction waves centered around
	the edges $x=\pm l$ of the initial distribution. These waves propagate inward
	the distribution with the linear sound velocity $c_s=1$ and collide each other
	at the center $x=0$ at the moment $t_c=l$. Thus, for $t\leq l$ the wave
	configuration consists of two rarefaction waves with unidirectional plasma
	flows and therefore they belong to the class of simple waves: in the
	right-propagating wave $r_+=\mathrm{const}$ and in the left propagating wave
	$r_-=\mathrm{const}$. The values of the constants are determined by the
	matching conditions at the edges propagating with sound velocities inward
	the region of quiescent plasma where $u=0$ and $\rho=1$, hence $r_{\pm}=0$.
	Thus, one of the Riemann invariant in the simple wave regions is known and the other
	must obey the remaining equation (\ref{t5-20.4}). In our case with sharp
	edges of the initial distribution its solution must be self-similar and
	depend only on one of the variables: $(x-l)/t$ for the right-propagating
	wave and $(x+l)/t$ for the left-propagating wave. These solutions can be easily
	obtained from the resulting equations $v_-=(x-l)/t$ or $v_+=(x+l)/t$,
respectively, and we arrive at simple formulas
	\begin{equation} \label{simpleraref}
	\begin{split}
	\rho(x,t) &  = \exp{\left( \frac{x+l-t}{t} \right)},  \qquad -\infty<x\le t-l, \\
	\rho(x,t) &  = \exp{\left( -\frac{x-l+t}{t} \right)},  \qquad l-t\le x<\infty.
	\end{split}
	\end{equation}
	In the case of isothermic equation of state the boundary with the vacuum
	disappears instantly and this means ``infinite'' velocity $u=x/t\pm1$ of the
	plasma flow at its tails. Such a non-physical behavior is a consequence of
the	supposition that the thermal equilibrium of electrons is maintained permanently
	due to their ``infinitely high'' temperature. Thus, the theory breaks down
	for the most energetic ions and needs some modification (see, e.g.,
	\cite{mora-03}). However, the number of such ions is exponentially small,
	$\rho\propto\exp(\mp x/t)$, and for the bulk of the density distribution the
	hydrodynamic approach remains accurate enough.
	
	\begin{figure}[t]
		\begin{center}
			\includegraphics[width=7cm]{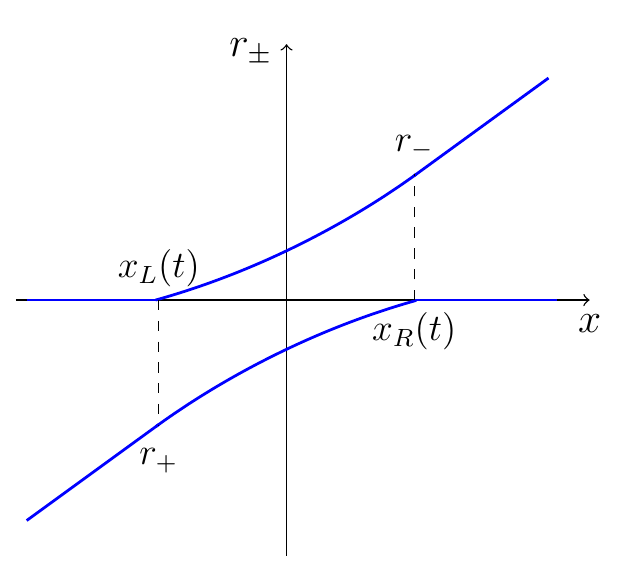}
			\caption{Dependence of the Riemann invariants $r_{\pm}$ on the space coordinate $x$
				at the fixed moment of time $t>l$.
			}
			\label{fig1}
		\end{center}
	\end{figure}

	After the moment $t=l$ of the collision of the rarefaction waves, a new
	region appears between them and in this region both Riemann invariants are
	changing with time; see Fig.~\ref{fig1}.
	Hence, we have to find such a solution of Eqs.~(\ref{t5-20.4})
	which matches at its edges $x_R(t),x_L(t)$ with the rarefaction wave
	solutions (\ref{simpleraref}).

	\section{Hodograph transform}
	
	In the region of the general solution both Riemann invariants $r_{\pm}=r_{\pm}(x,t)$
	are functions of $x$ and $t$ and they obey the nonlinear equations (\ref{t5-20.4}).
	These functions define mapping from the $(x,t)$-plane to the {\it hodograph} plane
	$(r_+,r_-)$. If this mapping can be inverted and we can consider $(r_+,r_-)$ as
	independent variables and $(x,t)$ as functions of them, then after this {\it hodograph
	transform} equations for the functions $x=x(r_+,r_-),t=t(r_+,r_-)$ become linear
	(see, e.g., \cite{LL-6,kamch-2000})
	\begin{equation}\label{HodTransEq}
	\begin{split}
	& \frac{\partial x}{\partial r_-}-v_{+}(r_{-},r_{+})\frac{\partial t}{\partial r_-}  = 0, \\
	& \frac{\partial x}{\partial r_+}-v_{-}(r_{-},r_{+})\frac{\partial t}{\partial r_+}  = 0.
	\end{split}
	\end{equation}
	If we look for the solution of this system in the form
	\begin{equation}\label{HodTransEqSol}
	\begin{split}
	x-v_{+}(r_{-},r_{+})t & = w_+(r_-,r_+), \\
	x-v_{-}(r_{-},r_{+})t & = w_-(r_-,r_+),
	\end{split}
	\end{equation}
	then, generally speaking, the functions $w_{\pm}$ must satisfy the Tsarev equations
	\cite{Tsa91}
	\begin{equation}\label{Tsarev}
	\begin{split}
	\frac1{w_+-w_-}\frac{\prt w_+}{\prt r_-}=
	\frac1{v_+-v_-}\frac{\prt v_+}{\prt r_-}, \\
	\frac1{w_+-w_-}\frac{\prt w_-}{\prt r_+}=
	\frac1{v_+-v_-}\frac{\prt v_-}{\prt r_+}.
	\end{split}
	\end{equation}
	In our case the right-hand sides of these equations are equal to $1/4$. Consequently,
	${\prt w_+}/{\prt r_-}={\prt w_-}/{\prt r_+}$, so we can introduce the potential $W$
	according to $w_{\pm}=\prt W/\prt r_{\pm}$, and then any of the Tsarev
	equations reduces to the Euler-Poisson equation for $W$:
	\begin{equation}\label{58-3}
	\frac{\prt^2W}{\prt r_+\prt r_-}-\frac{1}{4}\left(\frac{\prt W}{\prt r_+}
	-\frac{\prt W}{\prt r_-}\right)=0.
	\end{equation}
	At the boundary $x_R(t)$ with the right rarefaction wave we have $r_+=0,v_-=r_-/2-1=(x-l)/t$,
	hence the left-hand side of the first equation (\ref{HodTransEqSol}) equals to $-l$, and in
	a similar way the left-hand side of the second equation equals to $l$. Thus, we obtain the
	boundary conditions for the function $W(r_+,r_-)$ in the hodograph plane:
	\begin{equation}\label{eq20}
	\begin{split}
	& \frac{\prt W}{\prt r_-}=l\quad\text{at}\quad r_+=0,\\
	& \frac{\prt W}{\prt r_+}=-l\quad\text{at}\quad r_-=0.
	\end{split}
	\end{equation}
	The potential function $W$ is defined up to an additive constant, so we can fix it by
	the condition $W(0,0)=0$ and then Eqs.~(\ref{eq20}) give
	\begin{equation}\label{9-27}
	\begin{split}
	&  W(0,r_-)=l r_-\quad\text{along}\quad r_+=0,\\
	&  W(r_+,0)=-l r_+\quad\text{along}\quad r_-=0.
	\end{split}
	\end{equation}

	\section{Solution by the Riemann method}
	
	The Euler-Poisson equation (\ref{58-3}) with the boundary conditions (\ref{9-27}) can
	be solved by the Riemann method \cite{riemann,CH-89,Som64}. Since exposition of this method
	applied to similar problem has already been given in Refs.~\cite{ik-19,kamch-19},
	we shall not go into details here and just formulate the main principles.
	
	\begin{figure}[t]
		\begin{center}
			\includegraphics[width=7cm]{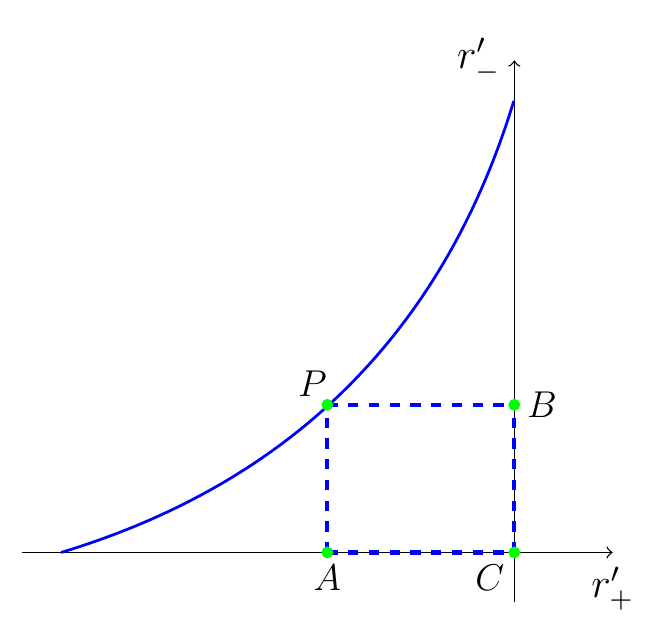}
			\caption{Contour $\mathcal{C}$ in the hodograph plane $(r_+',r_-')$ used in
				the Riemann method.
			}
			\label{fig2}
		\end{center}
	\end{figure}

\begin{figure}[t]
		\centering
		\includegraphics[width=0.45\textwidth]{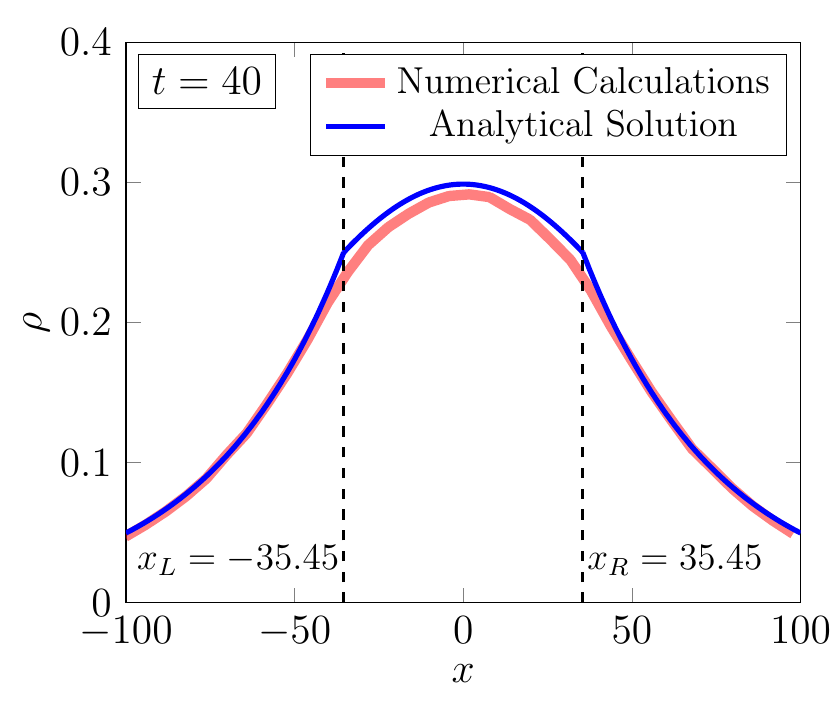}
		\caption{Distribution of the density $\rho(x)$ at
			$t = 40$ for $l = 20$ is shown by a thin (blue) line for analytical
			solution. Thick (red) line corresponds to the numerical solution. The symbols $x_R$
			and $x_L$ at vertical dashed lines indicate the boundaries between the
			general solution and the simple waves. }
		\label{fig3}
	\end{figure}

	Riemann showed \cite{riemann} that if we wish to find the value of $W$ at the point $P=(r_+,r_-)$
	in the hodograph plane (see Fig.~\ref{fig2}), then we should draw in this plane
	with coordinates $(r_+',r_-')$ two lines $PA$ $(r_+'=r_+=\mathrm{const})$ and
	$PB$ $(r_-'=r_-=\mathrm{const})$, which, together with $AO$ and $OB$ with known
	functions $W(r_+',0)$ and $W(0,r_-')$ along them, form a closed contour $\mathcal{C}=PAOB$.
	Since $(r_+,r_-)$ denote the coordinates of the ``observation point'' $P$, we have
	introduced here the notation $(r_+',r_-')$ for the coordinates in the hodograph plane.
	Following Riemann \cite{riemann}, we define in this plane such a vector $(V,U)$, that the integral
	$\int_{\mathcal{C}}(Vdr_+'+Udr_-')=0$ vanishes. The components $(V,U)$ depend both
	on the function $W$, which satisfies Eq.~(\ref{58-3}), as well as on another function
	$R=R(r_+',r_-';r_+,r_-)$ which must satisfy the condition of vanishing of the above
	integral. Riemann found that to this end one has to impose the following conditions:
	first, the function $R$ must satisfy the conjugate equation which in our case has the form
	\begin{equation}\label{11-28}
	\frac{\prt^2R}{\prt r_+'\prt r_-'}+\frac1{4}
	\left(\frac{\prt R}{\prt r_+'}-\frac{\prt R}{\prt r_-'}\right)=0;
	\end{equation}
	second, it must satisfy the boundary conditions
	\begin{equation}\label{11-29}
	\begin{split}
	& \frac{\prt R}{\prt r_+'}-\frac14{R}=0\quad\text{along}\quad PB;\\
	& \frac{\prt R}{\prt r_-'}+\frac14{R}=0\quad\text{along}\quad PA,
	\end{split}
	\end{equation}
	third, we fix its value at the point $P$,
	\begin{equation}\label{11-30}
	R(r_+,r_-;r_+,r_-)=1.
	\end{equation}
	If such a function $R$ is found, then the value of $W$ at the point $P=(r_+,r_-)$
	is given by the expression
	\begin{equation}\label{11-31}
\begin{split}
	W(P)&=\frac12(RW)_A+\frac12(RW)_B\\
&+\int_A^OVdr_+' +\int_O^BUdr_-',
\end{split}
	\end{equation}
	where $A=(r_+,0)$, $B=(0,r_-)$, $O=(0,0)$ and $U$, $V$ are expressed in terms of
the	boundary conditions (\ref{9-27}) by the formulas
	\begin{equation}\label{12-32}
	\begin{split}
	& U=\frac12\left({W}\frac{\prt R}{\prt r_+'}-R\frac{\prt {W}}{\prt r_+'}\right)
	-\frac14 WR,\\
	& V=\frac12\left(R\frac{\prt {W}}{\prt r_-'}-{W}\frac{\prt R}{\prt r_-'}\right)
	-\frac14 WR.
	\end{split}
	\end{equation}
	Thus, to get the solution, we have to find the Riemann functions $R$. Fortunately, for a gas
	with isothermal equation of state it was found by Riemann himself \cite{riemann} 
and in our notation it can be expressed in the form
	\begin{equation}\label{rim}
	\begin{split}
	R & = \exp{\left[ \frac14 (r_+-\xi-r_-+\eta) \right]}\\
	& \times
	I_0\left(\frac12 \sqrt{(r_+-\xi)(\eta-r_-)}\right),
	\end{split}
	\end{equation}
	where $I_0$ is the Bessel function of complex argument
	(see, e.g., \cite{WW-1927}).
	Substitution of Eq.~(\ref{9-27}) into Eq.~(\ref{11-31})
	followed by integration by parts with account of (\ref{12-32}) yields
	\begin{equation}\label{eq26}
	\begin{split}
	&W(r_+,r_-)=l\exp{\left(-\frac{r_+-r_-}{4}\right)} \\
	&\times \left\{ \int_{r_+}^{0}\left(1+\frac{r}{4}\right)
	\exp{\left(\frac{r}{4}\right)}I_0\left(\frac{1}{2}\sqrt{r_-(r-r_+)}\right)dr  \right. \\
	&+\left. \int_{0}^{r_-}\left(1-\frac{r}{4}\right)\exp{\left(-\frac{r}{4}\right)}
	I_0\left(\frac{1}{2}\sqrt{-r_+(r_--r)}\right)dr \right\}.
	\end{split}
	\end{equation}
	Once the function $W(r_+,r_-)$ is known, the dependence of $r_+$ and $r_-$
	on $x$ and $t$ is implicitly given by the formulas (\ref{HodTransEqSol}).
	Then substitution of these functions into Eqs.~(\ref{t5-22.16}) yields
	the distributions of the physical parameters of the plasma flow.
	We compare in Fig.~\ref{fig3} the analytical results (blue thin line)
	for the distribution of the density with the exact numerical solution
	(red thick line) of the system (\ref{PlasmaEq}). As we see,
	the hydrodynamic approximation agrees very well with the
	numerical solution.
	
	\section{Limiting cases}
	
	The solution obtained above provides the full description of plasma evolution
	during its expansion, but the formulas are quite complicated and therefore
	it is of considerable interest to obtain simpler results for some characteristic
	parameters of the flow. It is remarkable that the function $t=t(r_+,r_-)$ in the
	hodograph representation can be obtained from the system (\ref{HodTransEqSol})
	in a more direct way (see, e.g., \cite{gar-07}). Eliminating $x$ from this system,
	we arrive immediately at the Euler-Poisson equation for $t(r_+, r_-)$:
	\begin{equation}\label{tEP}
	\begin{split}
	\frac{\prt^2t}{\prt r_+\prt r_-}-\frac{1}{4}\left(\frac{\prt t}{\prt r_+}
	-\frac{\prt t}{\prt r_-}\right)=0,
	\end{split}
	\end{equation}
	which coincides with Eq.~(\ref{58-3}) for $W$. Therefore, its solution symmetric
	with respect to the transformation $r_+ \rightarrow -r_-$, $r_- \rightarrow -r_+$,
	and satisfying the condition $t (0,0) = l$ can also be expressed in terms of the
	Bessel function:
	\begin{equation}\label{tequation}
	\begin{split}
	t(r_+,r_-) = l \exp{\left(-\frac{r_+-r_-}{4}\right)}
	I_0\left(\frac{1}{2}\sqrt{-r_+r_-}\right).
	\end{split}
	\end{equation}
	At the right boundary with the rarefaction wave, the invariant $r_+$ vanishes and
	in a similar way the invariant $r_-$ vanishes at the boundary with the left
	rarefaction wave (see Fig.~\ref{fig1}). Then, as follows from Eq.~(\ref{tequation}),
	the Riemann invariants in the general solution change in the intervals
	\begin{equation}\label{}
	\begin{split}
	-4\ln{\left(\frac{t}{l}\right)}\le r_+ \le 0, \quad 0\le r_- \le 4\ln{\left(\frac{t}{l}\right)},
	\end{split}
	\end{equation}
	for a fixed value of $t>l$. At the right boundary $x_R(t)$ with $r_+=0$ the function
	(\ref{tequation}) reduces to $t(0,r_-)=le^{r_-/4}$, that is $r_-=4\ln(t/l)$ here.
	The right rarefaction wave solution at its boundary with the general solution is
	given by $x_R - (r_- / 2 - 1) t(0,r_-) = l$ and substitution of $t(0,r_-)$ and $r_-$
	yields the law of motion of the right boundary:
	\begin{equation}\label{k1}
	\begin{split}
	x_R(t) = l + 2t \ln{\left(\frac{t}{l}\right)} - t, \quad t\geq l.
	\end{split}
	\end{equation}
	Formula for the law of motion of the left boundary $x_L (t)$,
	due to the symmetry of the problem, differs from this only by the sign.
	
		\begin{figure}[t]
		\centering
		\includegraphics[width=0.45\textwidth]{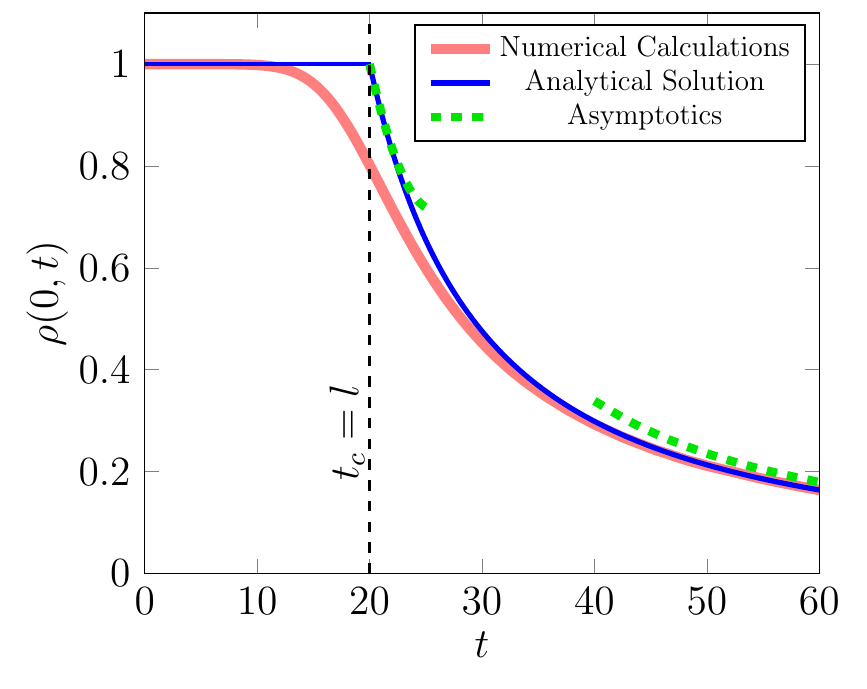}
		\caption{Dependence of the plasma density $\rho(0,t)$ at the center of the wave
			$x=0$ on time $t$ for $l = 20$.
			Numerical solution of the system (\ref{PlasmaEq}) is shown by a thick solid (red) line and analytical
			approximation by a thin solid (blue) line. Green dashed lines represent
			the asymptotic formulas (\ref{k3}) and (\ref{k4}).
			The vertical (black) dashed line shows the moment $t_c=l$
			when two rarefaction waves collide with each other
			at the center $x=0$.
			A small difference between the numerical and analytical results near the point $t_c$ arises 
due to dispersive effects, which are neglected in our hydrodynamic approximation.}
		\label{fig4}
	\end{figure}
	
	At the center $x = 0$ of the wave configuration we have $u = 0$, i.e., $r=r_- =-r_+$,
	and (\ref{tequation}) reduces to
	\begin{equation}\label{k2}
	\begin{split}
	t=t(-r,r) = l e^{r/2} I_0\left(\frac{r}{2}\right) =
	\frac{l}{\sqrt{\rho}} I_0\left(-\frac{1}{2}\ln{\rho}\right).
	\end{split}
	\end{equation}
	This equation determines implicitly the dependence of the plasma density $\rho$ on $t$
	at the distribution center.	Just after the moment of collision of two rarefaction waves,
	when $t-l\ll l$, we obtain
	\begin{equation}\label{k3}
	\begin{split}
	\rho(0,t) \approx 1-2\left(\frac{t}{l}-1\right)+\frac{7}{2}\left(\frac{t}{l}-1\right)^2,
	\quad \frac{t}{l}-1 \ll 1.
	\end{split}
	\end{equation}
	For asymptotically large time $t\gg l$ the density at the center $\rho(0,t)$
	is small and we can use the asymptotic formula for the Bessel function,
	$$
	I_0(z)\approx\frac{e^z}{\sqrt{2\pi z}},\quad z\gg1,
	$$
	which yields
	\begin{equation}\label{k4}
	\begin{split}
	\rho(0,t)=e^{-r}\approx \frac{l}{t\sqrt{\pi r}}\approx \frac{l}{t \sqrt{\ln{t/l}}}, \qquad t\gg l,
	\end{split}
	\end{equation}
	with logarithmic accuracy. All these formulas are confirmed by numerical solution
	of the system (\ref{PlasmaEq}) (see Fig.~\ref{fig4}).
	
	\section{Conclusion} In this Letter we have found exact analytical solution for the
	process of expansion of a slab of two-temperature plasma in hydrodynamic approximation.
	Comparison with the numerical solution of equations with account of dispersion effects,
	that is finite value of the Debye length, shows that this is a quite good approximation
	almost everywhere except for the regions of a very fast flow close to the boundaries
	of the plasma with a vacuum where the density is small. Thus, the obtained here solution
	allows one to get estimates of parameters of the plasma in the bulk of the wave
	configuration.

\end{document}